\documentclass{mem}
\usepackage{natbib}\usepackage{txfonts}\usepackage{balance}
\usepackage{graphicx}
\usepackage[a4paper,breaklinks,dvipdfm]{hyperref}
\idline{75}{1001}
\begin{document}
\def\teff{$T\rm_{eff }$}
\def\kms{$\mathrm {km s}^{-1}$}

\title{
Low-Mass X-ray Binaries in Globular Clusters: Puzzles and Solutions
}

   \subtitle{}

\author{
N. \,Ivanova\inst{1} 
          }

  \offprints{N. Ivanova}

\institute{
Department of Physics, University of Alberta, 11322 - 89 Avenue,
Edmonton T6G 2G7, AB, Canada. 
\email{nata.ivanova@ualberta.ca}
}

\authorrunning{Ivanova}

\titlerunning{LMXBs in Globular Clusters: Puzzles and Solutions}

\abstract{
In dense stellar systems, dynamical interactions between objects lead
to frequent formation of exotic stellar objects, unusual binaries,
and systems of higher multiplicity.
They are especially important for the formation of low mass X-ray binaries (LMXBs),
which are not only formed 100 times more efficiently than in the field,
but also have a puzzling dependence on metallicity.
In this contribution we review how compact objects are formed and retained,
the mechanisms of dynamical formation and the specifics 
of the evolution of mass-transferring binaries 
with neutron stars and black holes in globular clusters --
those two kinds of compact objects have different favored paths to become luminous in X-ray.
We describe how stellar evolution affects ostensibly purely dynamical formation, producing 
the observed metallicity dependence for LMXBs. We also discuss 
the next puzzle to be solved on our journey to understand the link between LMXBs and millisecond pulsars formation.

\keywords{ globular clusters: general  -- stellar dynamics -- X-rays -- binaries: close -- pulsars: general -- stars: neutron 
}
}
\maketitle{}

\section{Introduction}

Low-mass X-ray binaries (LMXBs) are Roche-lobe overflow (RLOF) binary stars consisting of a neutron star (NSs) or a black hole (BH) accretor and of a low-mass ($\la 1 M_\odot$) 
donor which can be a main-sequence (MS) star, a white dwarf (WD) or a red giant (RG). Globular clusters (GCs) are famous for  the number of LMXBs they contain per stellar mass as this exceeds the corresponding field LMXBs density by two orders of magnitude. This overabundance has been commonly interpreted as to be due to ability of stars in GCs to form LMXBs via dynamical encounters. Dynamical formation depends on the number of participants (the retention problem), the path how a compact object had acquired a companion (dynamical formation channels), and how a newly formed binary can proceed to, and evolve through, the mass-transfer; this affects the appearance in X-ray and the observed metallicity dependence. The success of some dynamical formation channels and the failure of others determines the population of formed LMXBs, providing a more solid link between the theory of their formation and the observed dependences. The final test in justification of theory of dynamical formation of LMXBs lies in the analysis of their progenies - millisecond pulsars (MSPs).

\section{Retention of compact objects}
 
\subsection{Neutron Stars}

NSs are usually formed via collapse of the core of a massive star during a supernova (we will refer to this channel of NS formation as to a core-collapse, CC). The mass range in single stars to form a CC NSs is $\sim 8-21 M_\odot $ for solar metallicity Z=0.02 and $\sim 7-19 M_\odot $ for Z=0.001. CC channel is an NS formation path which is accompanied by natal kicks that have a very high mean velocity, $\sim 400$ km/s  \citep{2005MNRAS.360..974H}. 
Discovering natal kicks immediately posed the problem: how NSs can be retained in GCs, where the escape velocities $v_{\rm esc}$ are several dozen of km/s. 
Indeed, a ``typical'' GC of $2\times 10^5 M\odot$,  with $v_{\rm esc}=40$km/s, will be able to retain only one NS if its stellar population was entirely consisting of single stars,  and only 15 if all stars were initially in binaries \citep{iva2008}. 
This is in direct contradiction with observations, where we know GCs that contain 
several dozen of identified radio MSPs  \citep{2012arXiv1210.3984F},
 and some are estimated to have more than a 100 \citep[e.g., Terzan 5,][]{2011MNRAS.418..477B}.

An alternative formation channel of NSs is via electron captures which start when a degenerate ONeMg core  reaches a critical density  -- this is known as electron capture supernova \citep[ECS,][]{1980PASJ...32..303M}.
ECS can occur during normal stellar evolution when a degenerate ONeMg core is  $\sim 1.36 M_\odot$ (evolutionary-induced collapse, EIC). 
The critical condition can be reached in a very narrow range of initial masses of single stars, where the exact values of the mass range depend on metallicity as well as on underlying assumptions how mixing and convective overshooting should be treated, and hence vary between the stellar codes. Being in a binary helps. First of all, the mass range of stars where EIC can proceed is larger, due to mass exchanges \citep{podsi04}. Secondly,  a degenerate ONeMg core with the critical central density could also be formed as a result of an accretion on a WD (accretion-induced collapse, AIC), or be formed as a result of either a merger (in GCs, also due to a collision) of two WDs (merger-induced collapse, MIC). 

In a field population of single stars, from 10\% to 15\% of all NSs ever formed can be formed via EIC \citep{iva2008}. Fraction of AIC NSs and MIC NSs is rather insignificant, in case of 100\% primordial binaries they provide ten times fewer NSs than EIC. However, since ECS NSs are born with much lower natal kicks \citep{2006A&A...450..345K}, a significant fraction of them can be retained in GCs. The mentioned above ``typical'' GC can retain as many as 200-300 NSs, 47 Tuc cluster could retain up to a 1000 of NSs \citep{iva2008}. As a result,  in GCs, ECS NSs are expected to exceed CC NSs by 30-200 times, providing enough NSs to explain the observations. Since ECS NSs are expected to be at a low-mass range, $1.22-1.27 M_\odot$, future mesaurements of masses of NSs in GCs can verify this hypothesis on the prevailance of ECS formation path for NSs in GCs.
 
\subsection{Black Holes}

BHs retention in GCs is less affected by natal kicks than the retention of NSs as many BHs are formed with kicks lowered through fallback \citep{bel06}. It can be estimated, that, depending on metallicity, each $150-200 M_\odot$ of stellar mass in a current ``aged'' globular cluster had produced a BH in the past (assuming no drastic mass loss event like tidal stripping occurred to that GCs). 40-50\% of the formed BHs would be retained after their formation in a relatively massive GCS with $v_{\rm esc}=50$ km/s \citep{bel06}. 

It was widely anticipated in the past that BHs could be effectively removed by another mechanism -- Spitzer instability would lead to a formation of a sub-cluster of BHs, decoupling BHs from the rest of GC stellar population. This sub-cluster is much denser than the rest of the GC core. Via strong interactions between each other, BHs quickly evaporate from the GCs, presumably leaving at the current epoch only one single or a binary pair BH \citep{kal04}. 
Detailed numerical calculations of such BH sub-clusters however showed that  in massive clusters  up to $\sim 20\%$ 
of the BHs may remain; and these BH sub-clusters do not reach equipartition \citep{2006ApJ...637..937O}.
Further Monte Carlo studies of a whole GC showed that  up to $25\%$ of initially retained BHs 
not only remained, but also  participated in interactions with other stars \citep{2010MNRAS.407.1946D}. Most recent studies claimed that the remaining fraction could be up to a half of initially retained BHs which form no BH sub-cluster at all \citep{2012arXiv1211.3372M}.

\section{Dynamical formation channels}

In GCs, most of NSs or BHs lose their primordial binary companion either before, 
or immediately on their formation, unless it was a NS formed via AIC. 
Compact objects then can acquire a new companion via exchange into a binary (BE, binary exchange), 
a physical collision (PC) with a giant or via a tidal capture (TC). 
The formation of a binary with a NS or a BH does not imply that this binary will become an LMXB. 
If a formed binary can become an LMXB within the Hubble time while evolved in isolation, we call it as a {\it direct} LMXB formation.

\subsection{LMXBs with NS accretors}

TCs were found to be the  least efficient mechanism of LMXBs formation in GCs; when succeed, 
they form predominantly LMXBs with MS donors \citep[this and further 
statistics in this subsection is taken from simulations described in ][]{iva2008}. 
PCs lead to a formation of NS-WD binaries \citep{iva05}, which, if become LMXBs, are known as ultra-compact X-ray binaries (UCXBs).
Most of TCs and PCs result in a direct NS-LMXB formation.
NSs which were formed via AIC, in most cases, would become LMXBs soon after NS formation as their orbit only slightly expanded;
donors in such LMXBs remain the same as at the moment of AIC and usually are MSs. 

BEs can form a binary with a NS and any type of a companion, however they usually do not lead to a  direct formation of NS-LMXBs.
To become an LMXB, a post-exchange binary would have to be altered further, via subsequent dynamical encounters. 
This usually occurs as a result of the process we term as ``eccentricity pumping''.
Eccentricity pumping is a result of a sequence 
of fly-by encounters which gradually boost eccentricity. 
A fly-by encounter is an encounter that occur at a distance large enough so it does not affect 
significantly a binary separation during one individual encounter, neither is leads to a companion exchange. 
The fly-by encounters are generally known to gradually harden a binary, 
decreasing its orbital separation, if that binary was initially hard \citep{1975MNRAS.173..729H}.
However, for binaries with NS, the more important role in a successful NS-LMXB formation 
was identified to be due to the eccentricity pumping.

BEs predominantly lead to a formation of LMXBs with MS and RG donors. 
Formation of UCXBs via BEs, while less productive than LMXBs formation with MS and RG donors, is still non-negligible, 
and produces just few times less UCXBs that PCs. 
We note that post-exchange NS binaries,  as well as post-PC NS binaries, are eccentric at the start of RLOF.

NS binaries can also form hierarchically stable triples through encounters with binaries. 
E.g., in a cluster similar to 47 Tuc, each NS binary has a 5\% chance to form a triple in a Gyr \citep{ivatriples}. 
A third of these triples is affected by a Kozai mechanism. Kozai mechanism,
if coupled with tidal friction, drives the inner binary to merge or RLOF before next interaction \citep{2006Ap&SS.304...75E}.
In numerical simulations, about a half of binary MSPs (bMSPs) were in triples at some point in their past!

\subsection{LMXBs with BH accretors}

Observationally implied formation rate of BH-LMXBs is very low: 
indeed, only one potential BH-LMXB is present at the time per $1-3\times 10^9 M_\odot$ in GCs \citep{iva10}. 
Hence theoretical studies of the formation of BH-LMXBs using a self-consistent 
dynamical code to model a complete GC are computationally unfeasible. 
To obtain statistics and rates, \cite{iva10} used a simplified approach, where the formation of BH-LMXBs was studied as per a BH
via separate formation channels. The most reliably identified at the time GC BH-LMXB, 
had low H$\alpha$/[OIII] ratio in its spectra and strong, broad OIII lines \citep{2008ApJ...683L.139Z}, signaling about likely WD donor. Hence only the formation of BH-WD LMXBs was studied in details.

It is hard to have a direct formation of BH-LMXBs: 
most of PCs lead to a formation of too wide binaries\citep{iva10}, AIC formation is not applicable,
BEs form too wide system \citep{kal04}.
Indeed, \cite{iva10} found that the formation of a BH-LMXB is necessarily a multi-step one. 
First step is the formation of a ``seed'' BH-WD binary. 
This occurs either via BE or a PC. Such a binary, with the rare 
exception of a very small fraction post-PC BH-WD binaries, can not become an LMXBs if evolved in isolation.

Post-PC BH-WD binaries however can be {\it transformed} into future BH-LMXBs via triple-induced mass transfer (TIMT).
For BH-WD binaries, the rate of formation of an hierarchically stable triple in encounters with binaries was found
 to exceed the rate of strong encounters with single stars, if the binary fraction is about a few per cent.  
A fraction of these hierarchically stable triples would be formed with high enough inclination to be affected by a Kozai mechanism. 
For the Kozai BH triples, a time-scale for reaching the maximum eccentricity was found to be 
shorter than  a time-scale for the next strong dynamical encounter. Hence, once formed, a seed BH-WD binary with 
a binary separation $\le 80R_\odot$ can be quickly brought to the regime where gravitational wave radiation can shrink the binary to RLOF within the Hubble time.

Most of post-BE BH-WD binaries are however too wide to be trasformated by TIMT. They need to be hardened, and this usually occurs as a result of many encounters; evolution though a hardening sequence can take up to a Gyr. Most of initially formed post-BE BH-WD binaries get destroyed along this sequence, but a few per cent of them survive until their binary separation is short enough for TIMT. (Hardening) $\rightarrow$ TIMT $\rightarrow$ gravitation wave radiation $\rightarrow$ UCXB sequence was shown to provide enough BH LMXBs to explain the observed number of BH-LMXBs (this required that at least 1\% of initially formed BHs is retained in a GC).
It is remarkable that recent observations showed that one of GC BH-LMXB is likely in a triple system \citep{2010MNRAS.409L..84M}.

Recently, two accreting BH-candidate binaries were detected in radio in M22 \citep{2012Natur.490...71S}. 
The potential companions for both of the objects, based on the identified optical counterparts,  are either low-mass MS stars, 
or WDs with the luminosities  below the detection limit of the used HST data.
The formation of BH-MS LMXBs  hence might be re-accessed in a future, with more thorough numerical studies
that employ a self-consistent method to emulate a whole GCs, and possibly with a better understanding of TCs by BHs.

\section{Metallicity dependence}

A metal-rich GC is $\sim 3$ times more likely to contain an LMXB than a metal-poor GC \citep{grin93}. This holds for Milky Way GCs as well as for extragalactic GCs.
For extragalactic sample this ratio also was found to hold across the range of X-ray luminosities, from $2\times10^{37}$ erg s$^{-1}$ to   $5\times10^{38}$ erg s$^{-1}$ \citep{2012arXiv1208.5952K}.
For even higher X-ray luminosity,
where the accretors are likely  BHs, this ratio is less certain, but still is above one, $2.5^{+0.9}_{-1.1}$.

Relatively low X-ray luminosities, $L_x\la 2\times 10^{37}$ erg s$^{-1}$, can be provided by LMXBs with a variety of donors,
however LMXBs with $L_X\ga 2\times10^{37}$ erg s$^{-1}$ are unlikely to be MS-LMXBs \citep{fra09}.
Indeed, in Milky Way GCs, only UCXBs are present at $L_x\ga 2\times 10^{37}$ erg s$^{-1}$.
Hence despite having the same ratio for Milky Way GCs and for extragalactic samples, 
the observed metallicity dependence must be analyzed separately -- different $L_X$ implies different donors!

Furthermore, in Milky Way GCs, there is no identified 
bright ($L_X\ga 10^{36}$ erg s$^{-1}$) LMXB with a MS donor in metal-poor GCs, while there are four bright MS-LMXBs in metal-rich GCs.  
This led to the idea that the formation of MS-LMXBs in metal-poor GCs is halted \cite{iva06}. In the MS-LMXBs the most effective  
mechanism of angular momentum removal is the magnetic braking, which operates only in stars with an outer convective zone.
Metal-poor MS stars lack an outer convective zone in a large mass range of low-mass  MS stars. 
If formed, a metal-poor MS-LMXBs is expected to be in quiescence for the most duration of the mass transfer.

The mechanism that prevents the formation of NS-MS LMXBs, 
however, can not explain the observed ratio of brighter,  $L_X\ga 2\times10^{37}$ erg s$^{-1}$,  
LMXBs in extragalactic GCs -- there the potential donors are likely RGs and WDs.
Formation of NS-RG donors is predominantly a direct LMXB formation via BEs, 
where the evolutionary expansion of a giant brings the system to the start of RLOF.
The formation rate of such systems depends on dynamical properties of a GC (also known as the stellar encounter rate),
as well as on the population of binaries with RGs  and the number of NSs.
Formation of NS-UCXBs in GCs is predominantly a direct LMXB formation via PC with a RG;  
post-PC binary shrinkage is solely due to angular momentum loss via gravitational wave radiation. 
The formation rate of UCXBs depends as well only a GC dynamical 
properties and the number of encounter participants, which are NSs and RGs.
The stellar encounter rate was found to be statistically identical in metal-rich and metal-poor GC samples \citep{2012arXiv1208.5952K}.
From our current understanding of NS formation and retention, it is expected 
that slightly more of NSs can be contained in a metal-poor GC than in a similar metal-rich GC 
-- the opposite trend to the metallicity dependence.  
Hence, the only remaining quantity that might be different is related to  RGs.

\cite{iva12} compared how the populations of RGs can differ 
between metal-rich and metal-poor GCs. 
It has been always appreciated that GCs of different metallicities 
would have different turn-off masses for their main sequences:  
metal-poor MS stars  evolve faster and hence metal-poor GCs have stars of lower mass at their MS turn-off. 
It appeared, that metal-poor stars live through RGs stage faster as well. 
Since RGs evolve faster, a metal-poor GCs would contain RGs that have a smaller mass range than RGs in a metal-rich GCs. 
Detailed evolutionary calculations showed that the number of stars that a metal-poor GCs can have in a RGs stage, 
assuming that GCs of different metallicities had the same initial mass function, 
is about 2 times smaller! 
The additional effects are: i) the lifetime of a RG-NS mass transferring system, which is shorter in metal-poor GCs; ii) the appearance of a RG-NS LMXBS -- metal-poor systems are predominantly transient while metal-rich are more likely to be persistent and iii) the mass of a RG, which is larger in a metal-poor RGs, boosts the collision rate with RGs and decreases the post-PC separations, resulting a direct LMXB formation for a larger number of encounters. Combined, all these effects  result in the about 3 times more frequent formation of UCXBs in metal-rich clusters than in metal-poor.

\section{Production of radio MSPs}

The anticipated ultimate fate of an NS-LMXB is to exhaust available donor's material, spin-up an accretor and, depending on how much mass was accreted, possibly appear as a MSP.
The comparison of the population of binary MSPs (orbital periods, companion masses, eccentricities) is therefor a crucial test for LMXB evolutionary theory.
As GCs are known to be LMXB factories, it has been expected that they would also contain a large number of MSPs. Extensive dedicated searches were performed specifically for GCs, and as a result 
most of identified anywhere radio MSPs (144 to-the-date) are located in GCs \citep{2012arXiv1210.3984F}, 
and 2/3 of them are in binaries. 

The comparison of bMSPs populations in simulations of GCs and in observations showed nice agreement for the  most of the formation channels: for BE (with a MSP being exchanged into a binary) as well as for post-mass transfer system with initial RG and MS donors. However, a striking disagreement  was found for the  theoretically expected progenies of UCXBs. Both simplified estimates and detailed numerical simulations of GCs were able to re-produce a number of UCXBs comparable to the observed number of UCXBs in GCs of different dynamical properties, however this did not occur for post-UCXBs bMSPs. 

The lifetime of an UCXB at $L_X>10^{36}$ erg s$^{-1}$ is only $\sim 10^8$ years. 
If the formation rate of UCXBs in the past was not drastically different from the current, then each currently present  
UCXB with $L_X>10^{36}$ erg s$^{-1}$  implies a hundred of UCXBS formed in the past; those UCXBs should have become bMSPs by now. 
Milky Way GCs contain about a dozen of such bright UCXBs at present, 
implying that there should be a thousand of post-UCXBs bMSPs.  
However, they are completely lacking in the observed sample of {\it radio} bMSPs in GCs, 
both by the number and by orbital period-donor mass distribution  \citep{iva2008}. 
 A similar lack of post-UCXBs radio bMSPs was identified at the same time for a field population of bMSPs \citep{del08}. 
We note that at least one post-UCXBs bMSP was detected in X-rays and $\gamma$-rays. \citep{2007ApJ...668L.147K}.
The reason for post-UCXBs MSPs to be radio-quiet is not yet understood and is a new puzzle posed by GCs for us.

\section{Conclusions}

\begin{itemize}

\item[$\bullet$] Population of NSs in GCs relies upon electron capture supernovae: retained ECS NSs  are expected to exceed CC NSs by far. 

\item[$\bullet$]  Detections of BH-LMXB in GCs indicated that BH population in GCs is non-negligible. 
This agrees with the results of most recent dynamical simulations of GCs.

 \item[$\bullet$]  LMXBs in different ranges of X-ray luminosity are formed via different dynamical formation channels. 
The metallicity dependance is mainly determined by red giants life-times for $L_X\ga 2\times 10^37$ erg s$^{-1}$,
but is also influenced by magnetic braking for a lower-range of bright LMXBs.

\item[$\bullet$]  Dynamical formation of hierarchically stable triples is a crucial process for LMXBs in GCs with both BHs and NSs accretors. 

\item[$\bullet$]  Mass transfer in UCXBs does not result in the formation of radio bMSPs.

\end{itemize}
 
\begin{acknowledgements}
NI aknowledges  support by NSERC Discovery Grant and the Canada Research Chairs program.

\end{acknowledgements}

\bibliographystyle{aa}

\begin{thebibliography}{}

\bibitem[Bagchi et al.(2011)]{2011MNRAS.418..477B} Bagchi, M., Lorimer, 
D.~R., \& Chennamangalam, J.\ 2011, \mnras, 418, 477 

\bibitem[Belczynski et al.(2006)]{bel06} Belczynski, K., 
Sadowski, A., Rasio, F.~A., \& Bulik, T.\ 2006, \apj, 650, 303 

\bibitem[Deloye(2008)]{del08} Deloye, C.~J.\ 2008, 40 Years 
of Pulsars: MSPs, Magnetars and More, 983, 501 



\bibitem[Downing et al.(2010)]{2010MNRAS.407.1946D} Downing, J.~M.~B., 
Benacquista, M.~J., Giersz, M., \& Spurzem, R.\ 2010, \mnras, 407, 1946 

\bibitem[Eggleton 
\& Kisseleva-Eggleton(2006)]{2006Ap&SS.304...75E} Eggleton, P.~P., \& Kisseleva-Eggleton, L.\ 2006, \apss, 304, 75 


%\bibitem[Fragos et al.(2008)]{fra08} Fragos, T., Kalogera, 
%V., Belczynski, K., et al.\ 2008, \apj, 683, 346 
 
\bibitem[Fragos et al.(2009)]{fra09} Fragos, T., Kalogera, 
V., Willems, B., et al.\ 2009, \apjl, 702, L143 



\bibitem[Freire(2012)]{2012arXiv1210.3984F} Freire, P.~C.~C.\ 2012, 
arXiv:1210.3984 

\bibitem[Grindlay(1993)]{grin93} Grindlay, J.~E.\ 1993, The 
Globular Cluster-Galaxy Connection, 48, 156 

\bibitem[Heggie(1975)]{1975MNRAS.173..729H} Heggie, D.~C.\ 1975, \mnras, 
173, 729 

\bibitem[Hobbs et al.(2005)]{2005MNRAS.360..974H} Hobbs, G., Lorimer, 
D.~R., Lyne, A.~G., \& Kramer, M.\ 2005, \mnras, 360, 974 

\bibitem[Ivanova et al.(2005)]{iva05} Ivanova, N., et al. \ 2005, \apjl, 621, L109 




\bibitem[Ivanova(2006)]{iva06} Ivanova, N.\ 2006, \apj, 636, 
979 

\bibitem[Ivanova et al.(2008)]{iva2008} Ivanova, N., Heinke, 
C.~O., Rasio, F.~A., Belczynski, K., 
\& Fregeau, J.~M.\ 2008, \mnras, 386, 553 


\bibitem[Ivanova(2008)]{ivatriples} Ivanova, N.\ 2008, Multiple 
Stars Across the H-R Diagram, 101 

\bibitem[Ivanova et al.(2010)]{iva10} Ivanova, N., 
Chaichenets, S., Fregeau, J., et al.\ 2010, \apj, 717, 948 


\bibitem[Ivanova et al.(2012)]{iva12} Ivanova, N., Fragos, 
T., Kim, D.-W., et al.\ 2012, \apjl, 760, L24 


\bibitem[Kalogera et al.(2004)]{kal04} Kalogera, V., King, 
A.~R., \& Rasio, F.~A.\ 2004, \apjl, 601, L171 

\bibitem[Kim et al.(2012)]{2012arXiv1208.5952K} Kim, D.-W., Fabbiano, G., 
Ivanova, N., et al.\ 2012, arXiv:1208.5952 



\bibitem[Kitaura et 
al.(2006)]{2006A&A...450..345K} Kitaura, F.~S., Janka, H.-T., \& Hillebrandt, W.\ 2006, \aap, 450, 345 

\bibitem[Krimm et al.(2007)]{2007ApJ...668L.147K} Krimm, H.~A., Markwardt, 
C.~B., Deloye, C.~J., et al.\ 2007, \apjl, 668, L147 

\bibitem[Maccarone et al.(2010)]{2010MNRAS.409L..84M} Maccarone, T.~J., 
Kundu, A., Zepf, S.~E., \& Rhode, K.~L.\ 2010, \mnras, 409, L84 

\bibitem[Miyaji et al.(1980)]{1980PASJ...32..303M} Miyaji, S., Nomoto, K., 
Yokoi, K., \& Sugimoto, D.\ 1980, \pasj, 32, 303 


\bibitem[Morscher et al.(2012)]{2012arXiv1211.3372M} Morscher, M., Umbreit, 
S., Farr, W.~M., \& Rasio, F.~A.\ 2012, arXiv:1211.3372 




\bibitem[O'Leary et al.(2006)]{2006ApJ...637..937O} O'Leary, R.~M., et al.
%Rasio, F.~A., Fregeau, J.~M., Ivanova, N., \& O'Shaughnessy, R.
\ 2006, \apj, 637, 937 




\bibitem[Podsiadlowski et al.(2004)]{podsi04} Podsiadlowski, 
P., Langer, N., Poelarends, A.~J.~T., et al.\ 2004, \apj, 612, 1044 




\bibitem[Strader et al.(2012)]{2012Natur.490...71S} Strader, J., et al.\
%Chomiuk, L., Maccarone, T.~J., Miller-Jones, J.~C.~A., \& Seth, A.~C.\ 
2012, \nat, 490, 71 

\bibitem[Verbunt 
\& Lewin(2006)]{gcbook} Verbunt, F., \& Lewin, W.~H.~G.\ 2006, Compact stellar X-ray sources, 341 


\bibitem[Zepf et al.(2008)]{2008ApJ...683L.139Z} Zepf, S.~E., Stern, D., 
Maccarone, T.~J., et al.\ 2008, \apjl, 683, L139 


\end{thebibliography}

\end{document}